\documentclass[10pt]{article}
\usepackage[square,authoryear]{natbib}

\RequirePackage{graphicx}
\RequirePackage{verbatim}
\RequirePackage{amssymb,amsmath,amsthm}

\RequirePackage{color}
\definecolor{dullmagenta}{RGB}{102,0,102}   

\PassOptionsToPackage{colorlinks
	   ,urlcolor=dullmagenta
           ,citecolor=dullmagenta
           ,linkcolor=dullmagenta
           ,pagecolor=white
 	   }{hyperref}
\IfFileExists{hyperref.sty}{\RequirePackage{hyperref}}%
 {\typeout{hyperref.sty is not loaded so you will not get active hyperlinks}%
 \providecommand{\href}[2]{#2}%
}

\def\intprod{\mathbin{\hbox to 6pt{ \vrule height0.4pt width5pt depth0pt
\kern-.4pt \vrule height6pt width0.4pt depth0pt\hss}}}

\makeatletter
 \@addtoreset{figure}{section}
 \def\thefigure{\thesection.\@arabic\c@figure}
 \def\fps@figure{h, t}
 \@addtoreset{equation}{section}

\def\editedby#1{\gdef\@editedby{#1}}
\def\@editedby{\@latex@error{No \noexpand\editedby given}\@ehc}
\editedby{}

\def\SetVersionDate#1{\gdef\@SetVersionDate{#1}}
\def\@SetVersionDate{\@latex@error{No \noexpand\SetVersionDate given}\@ehc}
\SetVersionDate{\today}
\gdef\VersionDate{\@SetVersionDate}

\newif\ifdraftflag@     \draftflag@false
\def\Draft{\global\draftflag@true
            \message{Draft Mode Activated}
  \def\@oddfoot{\footnotesize\upshape
       {\bf Version \VersionDate}\dotfill
        Edited by \@editedby \ :
      {Typeset on \today\ --\hhmm}
               }
  \def\@evenfoot{\footnotesize\upshape
       {\bf Version \VersionDate}\dotfill
        Edited by \@editedby \ :
      {Typeset on \today\ --\hhmm}
               }
\def\idxcolor{red}
\def\idxcolordbl{blue}
\def\noidxcolor{green}
}

\def\DraftOff{\global\draftflag@false}

\def\markbothsame#1{\markboth{#1}{#1}}
\def\ps@myheadings{
  \let\@mkboth\markboth
\ifdraftflag@
  \def\@oddfoot{\footnotesize\upshape
       {\bf Version \VersionDate}\dotfill
        Edited by \@editedby \ :
      {Typeset on \today\ --\hhmm}
               }
  \def\@evenfoot{\footnotesize\upshape
       {\bf Version \VersionDate}\dotfill
        Edited by \@editedby \ :
      {Typeset on \today\ --\hhmm}
               }
\else
  \let\@oddfoot\@empty\let\@evenfoot\@empty
\fi
\def\@oddhead{\bf\small\upshape\hfil\rightmark\hskip\tw@ em\thepage}
\def\@evenhead{\bf\small\upshape\thepage\hskip\tw@ em\leftmark\hfill}
  \def\chaptermark##1{\markbothsame
    {\ifnum\c@secnumdepth>\m@ne\@chapapp\ \thechapter. \ \fi##1}}
  \def\sectionmark##1{\markright{\ifnum\c@secnumdepth>\z@\thesection\ \fi
    ##1}}
  \def\subsectionmark##1{\markright{\ifnum\c@secnumdepth>\z@\thesubsection\ \fi
    ##1}}
}

\def\ps@empty{%
\def\@oddhead{\hfill\raise\headheight\@crosshairs}%
\let\@evenhead\@oddhead
\def\@evenfoot{}\let\@oddfoot\@evenfoot
  \ifdraftflag@
  \def\@oddfoot{\footnotesize\upshape
       {\bf Version \VersionDate}\dotfill
        Edited by \@editedby \ :
      {Typeset on \today\ --\hhmm}
               }
  \def\@evenfoot{\footnotesize\upshape
       {\bf Version \VersionDate}\dotfill
        Edited by \@editedby \ :
      {Typeset on \today\ --\hhmm}
               }
  \fi
}
\def\todayfr{\number\day\space
 \ifcase\month
 \or janvier\or f\'evrier\or mars\or avril\or mai\or juin
 \or juillet\or aout\or septembre\or octobre\or novembre\or d\'ecembre
 \fi
 \space\number\year}
\def\today{\number\day\space
    \ifcase\month
    \or January\or February\or March\or April\or May\or June
    \or July\or August\or September\or October\or November\or December
    \fi
    \space\number\year}
\def\Month{
    \ifcase\month
    \or January\or February\or March\or April\or May\or June
    \or July\or August\or September\or October\or November\or December
    \fi}

\def\MonthAndYear{
    \ifcase\month
    \or January\or February\or March\or April\or May\or June
    \or July\or August\or September\or October\or November\or December
    \fi
    \space\number\year}
\def\hhmm{
    \count1=\time
    \count2=\count1
    \divide \count1 by 60
    \count3=\count1
    \multiply \count1 by 60
    \advance\count2 by -\count1
    \number\count3 h%
    \ifnum\count2>9 \else 0\fi
    \number\count2
   }

\newcommand{\jemfootnote}{\null} 
\newif\ifAllowFootnote@     \AllowFootnote@false
\def\AllowFootnote{\global\AllowFootnote@true
            \message{Allow Footnotes Activated}
\renewcommand{\jemfootnote}{\footnote}
}
\newif\ifprtprecite@  \prtprecite@false
\def\PreCite#1{\gdef\@PreCite{#1}\ifprtprecite@\OutputPreCite\fi}
\def\@PreCite{\@latex@error{No \noexpand\PreCite given}\@ehc}
\gdef\OutputPreCite{\@PreCite}
\def\PrintPreCite{\global\prtprecite@true}
\def\eqref#1{(\ref{#1})}

\let\citejem\jemcite 
\let\cite\citejem

\expandafter\ifx\csname natexlab\endcsname\relax\fi

\let\NAT@citex@ori=\NAT@citex
\def\NAT@citex[#1][#2]#3{\NAT@fulltrue\NAT@citex@ori[#1][#2]{#3}}

\makeatother

\usepackage{epstopdf}
\usepackage{setspace}
\usepackage{bbm}

\usepackage{amscd}

\usepackage{caption2}

\DeclareGraphicsExtensions{.eps,.ps,.pdf}

\begin{document}

 \newtheorem{thm}{Theorem}[section]
 \newtheorem{cor}[thm]{Corollary}
 \newtheorem{lem}[thm]{Lemma}
 \newtheorem{prop}[thm]{Proposition}
 \newtheorem{defn}[thm]{Definition}
 \newtheorem{rem}[thm]{Remark}
 \numberwithin{equation}{section}

\title{\Large{\textbf{Estimating Terminal Velocity of Rough Cracks 
\\in the Framework of Discrete Fractal Fracture Mechanics
}}}

\author{Arash Yavari\thanks{School of Civil and Environmental Engineering,
  Georgia Institute of Technology, Atlanta, GA 30332. E-mail: arash.yavari@ce.gatech.edu.}
  \and Hamed Khezrzadeh \thanks{Department of Civil Engineering, Center of Excellence in Structures and Earthquake Engineering,
   Sharif University of Technology, P.O. Box 11155-9313, Tehran, Iran.}
}

\maketitle

\begin{abstract}
In this paper we first obtain the order of stress singularity for
a dynamically propagating self-affine fractal crack. We then show
that there is always an upper bound to roughness, i.e. a propagating fractal
crack reaches a terminal roughness. We then study the phenomenon of
reaching a terminal velocity. Assuming that propagation of a
fractal crack is discrete, we predict its terminal velocity using an asymptotic energy
balance argument. In particular, we show that the limiting crack
speed is a material-dependent fraction of the corresponding
Rayleigh wave speed.
\end{abstract}

\begin{description}
\item[Keywords:] Dynamic fracture, Fractal crack, Order of stress singularity, terminal velocity.
\end{description}


\section{Introduction}

A theoretical framework for including inertial effects during a
rapid crack growth was first proposed by \citet{Mott1948}, who
adopted the analysis of \citet{Griffith1921} as a starting point. The
idea is based on a simple addition of a kinetic energy term to the
expression for the total energy of the cracked system. According
to Mott's extension of Griffith's criterion, the requirement that
the system remains in thermodynamic equilibrium with its
surroundings as the crack extends leads to the following
expression in terms of the well-known fracture parameters:
$G-2\gamma=dT/da$, where $G$ is energy release rate, $\gamma$ is
specific surface energy, $T$ is kinetic energy density, and ``a"
is the characteristic length of the crack. Mott defined a domain
\emph{R} that receives stress-wave ``messages" from the crack tip
and then argued that the total kinetic energy can be written as
$T=\frac{1}{2}\rho v^2\int_{R}[(\partial u_x/\partial
a)^2+(\partial u_y/\partial a)^2]dxdy$. While Mott's analysis
lacks rigour, it is instructive in the way it highlights some of
the important features of a running crack without excessive
mathematical complication.

The first important contribution to the problem of a moving crack
with constant velocity was the work of \citet{Yoffe1951}. The Yoffe
problem consists of a mode I crack of fixed length traveling
through an elastic body at a constant speed under the action of
uniform remote tensile loading. \citet{Yoffe1951} obtained the
stress distribution near the tip of a rapidly propagating crack
in a plate of isotropic elastic medium. The result was that the
stresses depend on the crack tip velocity and reduce to
the solution of \citet{Inglis1913} when the velocity is zero.

\citet{Roberts1954} used Mott's extension of Griffith's criterion
to predict the limiting velocity of the crack extension. By
taking the boundary of the region $R$ to be a circle of radius
$r$ centered at the crack tip, they estimated $r\approx c_0t$,
where $c_0=\sqrt{E/\rho}$ is the longitudinal sound wave speed. They defined this cutoff region as the
border of the disturbed zone by the stress waves emanated from
the crack tip. Using this assumption and taking a stress field
similar to that of the static case they roughly estimated the
limiting crack velocity to be about $0.38c_0$ when
$\nu=0.25$. \citet{Steverding1970} studied the problem of
the response of cracks to stress pulses and found an equation of
motion for such cracks. They also obtained the limiting velocity of
crack extension caused by stress pulses by using asymptotic
solutions to be about $0.52c_R$, where $c_R$ is the Rayleigh wave
speed. There have also been some other efforts on finding the
equation of motion for dynamically propagating cracks.
\citet{Berry1960a,Berry1960b} was the first to find an equation
of motion for dynamic propagation of cracks. He found out that
the details of the motion of a crack are determined by the state of stress
at the point of fracture, and that the observed
critical stress is (infinitesimally) greater than that given by
the Griffith's criterion and is probably determined by the size of
the defect in the sample and the rate of straining. He also obtained
solutions for the equation of motion for fracture in
tension and fracture in cleavage in both constant force and
constant velocity cases. See \citet{Bouchbinder2010} for a recent review of dynamic fracture mechanics.

The inadequacy of the classical fracture mechanics theories in
problems such as predicting infinite strength for elastic
bodies without any cracks, for example, was the motivation for
some researchers to propose new failure theories.
\citet{Novozhilov1969} introduced a non-local stress criterion
and gave the condition of the brittle crack propagation in mode I
as $\sigma^*\equiv\langle \sigma_y(x)\rangle_0^{a_0}=\sigma_c$,
where $\langle . \rangle_0^{a_0}$ is spatial averaging over the
interval $[0,a_0]$, $\sigma_y(x)$ is the complete (not only
asymptotic) stress field around the crack tip $(x=0)$,
$\sigma_c$ is the ideal strength of the material, and $a_0$ is
the minimum admissible crack advance named by him a
\emph{fracture quantum}. According to Novozhilov this criterion
can be used only with the complete expression of the stress-field,
and not with its asymptotic form. However, the complete
expression is rarely known. Another restriction in Novozhilov's
approach was that the size of fracture quantum assumed to be the
atomic spacing. \citet{Pugno2004} introduced their so-called
quantized fracture mechanics (QFM) approach, which modified
Novozhilov's theory. In QFM, the restrictions of Novozhilov's
theory were removed and this made QFM a useful approach for
analysis of very short cracks (see also \citet{Krasovs'kyi2006,MorozovPetrov2002,CornettiPugnoCarpinteriTaylor2006,Leguillon2002}  for more related works). In their approach
the differentials in Griffith's criterion were replaced by
finite differences (see \citet{Wnuk2008} for a discussion). For
vanishing crack length, QFM predicts a finite ideal strength in
agreement with the prediction of \citet{Orowan1955}.

In most models in fracture mechanics cracks are assumed to be
smooth for mathematical convenience. However, in reality fracture
surfaces are rough and ``roughness" evolves in the process of crack propagation.
Fracture surfaces of many materials of interest are fractals, a
fact that has been experimentally established by many
researchers. A fractal dimension (roughness exponent) is not
enough to uniquely specify a fractal set and this is why all one
can hope for achieving having only a fractal dimension (roughness
exponent) is a qualitative analysis. Effects of fractality on
fracture characteristics of rough cracks have been investigated by
several groups in the past two decades (see \citet{Mosolov1991,Gold1991,Gold1992,Balankin1997,Borodich1997,Carpinteri1994}, \citet{Cherepanov1995,Xie1989,Xie1995,Yavari2000,Yavari2002a,Yavari2002,Yavari2002b,Wnuk2003,Wnuk2005,Wnuk2008,Wnuk2009}, and
references therein). Here our interest is to estimate the observed terminal
velocity of a rough crack propagating dynamically in an elastic
medium.

\citet{Wnuk2008} extended quantized (finite) fracture mechanics ideas for fractal cracks. They presented a modification of the classical theory of brittle fracture of solids by relating discrete nature of crack propagation to the fractal geometry of the crack. Their work is based on the idea of using an equivalent smooth blunt crack with a finite radius of curvature at its tips for a given fractal crack \citep{Wnuk2003, Wnuk2005}. By taking the radius of curvature of the equivalent blunt crack as a material property, they showed that fractal dimension of the crack trajectory is a monotonically increasing function of the nominal crack length. This result was an analytical demonstration of the mirror-mist-hackle phenomenon for rough cracks. Later they showed that assuming a cohesive zone ahead of a fractal crack, the size of the cohesive zone increases while the crack propagates \citep{Wnuk2009}.

To our best knowledge, the only contributions related to the dynamic fracture of fractal cracks are \citet{Xie1995} and \citet{Alves2005}. \citet{Xie1995} introduced a fractal kinking model of the crack extension path to describe irregular crack growth. Then by using the formula proposed by \citet{Freund1998} for calculating dynamic stress intensity factor for arbitrary crack tip motion he calculated the stress intensity factor for the assumed fractal crack path. He concluded that the reason for having terminal velocities lower than the Rayleigh wave speed is the fractality of the crack path. \citet{Alves2005} used a fractal model for rough crack surfaces in brittle materials. He tried to explain the effect of fractality of fracture surfaces
on the stable (quasi-static) and unstable (dynamic) fracture resistance. He concluded that fractal dimension has a strong influence on the rising of the R-curve in brittle materials. He also argued that the reason for having terminal velocities lower than Rayleigh wave speed is the roughness of the fracture
surfaces that makes the nominal (projected) and local crack tip velocities different.

\citet{Dulaney1960} modified Mott's analysis of energy balance
and showed that for a crack of initial and current lengths $c_0$
and $c$, respectively, $\dot{c}=v_L(1-c_0/c)$ compared to the
similar equation $\dot{c}=v_L(1-c_0/c)^{\frac{1}{2}}$ obtained by
Mott. Here $v_L$ is the terminal velocity and is proportional to
$\sqrt{E/\rho}$. They also carried out tests of terminal velocity
on PMMA specimens and showed that the measured velocity differs
only about ten percent from the predicted value by
\citet{Roberts1954}. Recently,
\citet{Chekunaev2008,Chekunaev2009} studied the terminal velocity
by replacing a mode I crack under pressure on its faces by considering
cohesive forces and replacing the crack by an equivalent
distribution of dislocations. They obtained simple expressions for the potential and kinetic energies of the
environment of the moving crack. They also obtained an expression
for an equivalent mass for the crack tip, i.e. a point mass that
has the same kinetic energy as the whole cracked system. Their
equivalent mass depends on a truncation radius $R$ in the form of
$\ln (2R/a)$, where $a$ is the half crack length. For a uniform
external pressure $p_0$ they showed that the crack tip speed can
be expressed as $v=v_L(1-a_{cr}/a)$, where $a_{cr}$ depends on
both mechanical properties and $p_0$. Their terminal velocity has
the form $v_L=g(v,R/a)\sqrt{E/\rho}$ that they approximate by
$v_L=\hat{g}(v)\sqrt{E/\rho}$, for functions $g$ and $\hat{g}$
that are given in \citep{Chekunaev2008,Chekunaev2009}. It should be mentioned that
their equivalent mass is positive for all $R/a>0$ and
$\nu\in(-1.0,0.5)$. Note that their equivalent mass is independent of the crack tip speed and hence kinetic energy is an increasing function of crack tip velocity.

This paper is organized as follows. In \S2 an asymptotic method will be used to determine the order of stress singularity for a dynamically propagating fractal crack. In \S3 dynamic propagation of a fractal crack is investigated. We first show that in the intermediate crack growth regime, i.e. after the initial phase of crack growth, roughness change is so small that a terminal roughness exponent can be assumed. The phenomenon of reaching a limiting speed is predicted using some simplifying assumptions. The predicted limiting crack speeds for different brittle amorphous materials are shown to have good agreement with the experimental results. Finally, conclusions are given in \S4.

\section{Order of stress singularity for a dynamically propagating fractal crack}

Consider a crack, whose tips are growing in opposite directions with equal velocities.\footnote{To avoid problems with stress wave reflections we can assume that the crack is semi-infinite. This assumption will not change anything in the following analysis.} The crack unloads some area of the body, and while propagating, size of the unloaded area will increase. The form of the stress field in the close vicinity of the crack tip is of interest as almost all fracture energy will be consumed through different processes in this zone. In the following, the asymptotic behavior of some important parameters that contribute to energy balance will be investigated. Here, we assume that the singularity of stress is of the form $r^{-\beta}$, where $r$ is distance from the crack tip. In the sequel we find an expression for the order of stress singularity $\beta$.

Several experimental observations confirm that energy consuming phenomena such as temperature rise, acoustic and phonon emissions, etc. occur in the close vicinity of the moving crack tip. There is one common aspect in all these phenomena; they all have a kinetic origin. To be more precise they are all results of fast movements or oscillations of the particles around the moving crack tip. This means that the velocity of the particles around the crack tip is of great importance. The particle velocity at a point depends on two main parameters: the crack tip speed and the strain at the point \citep{Freund1998}. Asymptotic behavior of particle velocity in the close vicinity of a smooth moving crack tip is similar to that of stress for all modes of
fracture. For the classic case we have $\dot{u}_i\sim r^{-\frac{1}{2}}$. The classic case is a limiting case of fractal model, which is when the roughness (Hurst) exponent ($H$) of a self-affine crack trajectory is equal to unity. In the case of a fractal crack, we know that $\dot{u}_i=f_i(v,K_I^f,r,E)$ and hence dimensional analysis tells us that
\begin{equation}
    \dot{u}_i= v \frac{K_I^f(t)}{r^{\beta}E}\Psi_{i}(\theta,v),
\end{equation}
where $\Psi_i$ is a dimensionless function and $v$ is short for $v_{\text{nominal}}$.

Asymptotic behavior of the true length of the crack trajectory is
also important. The first experimental study on fractal
characteristics of fracture surfaces was carried out by
\citet{Mandelbrot1984} who showed that the fracture surfaces of
steel are fractals. Since then many experimental investigations
have been done. For example, the investigations on the concrete
fracture surfaces by \citet{Saouma1990} and \citet{Saouma1994}
showed that the fracture surfaces of concrete are also fractals.
Based on the experimental observations, we assume that fracture
surfaces are self-affine fractals. The asymptotic behavior of the
true crack length growth $\Delta L$ is
\citep{Mandelbrot1985,Mandelbrot1986a,Mandelbrot1986b, Yavari2002}
\begin{equation}
    \Delta L\sim\ell^{\frac{1}{H}},
\end{equation}
where $\ell$ is some characteristic crack growth
length.\footnote{There are different definitions of fractal
dimension, e.g. box dimension ($D_B$), compass dimension ($D_C$),
and mass dimension ($D_M$). In the case of a self-similar
fractal, all of these dimensions have the same value, but this is
not the case for self-affine fractal sets. The local values of
the box dimension (using small boxes) and mass dimension (using
small radii) are both $2-H$. The compass dimension has a local
value $\frac{1}{H}$. For more details see
\citep{Mandelbrot1985,Mandelbrot1986a,Mandelbrot1986b}.}.

The changes of kinetic and strain energies of the body $\mathcal{R}$ when the crack length growth is $\Delta L$ are (summation over repeated indices is implied):
\begin{equation}
    \Delta T = \int_{\mathcal{R}}
    \frac{1}{2}\rho\,\dot{u}_i\,\dot{u}_i\,dA~~~~~\textrm{and}~~~~~\Delta U_e = \int_{\mathcal{R}}
    \frac{\sigma^2}{E}\,dA,
\end{equation}
where $dA$ is the area element. If it is assumed that the change of the strain and kinetic
energies is dominant in a small neighborhood $\mathcal{R}_s$
around the crack tip, in the above relations $\mathcal{R}$ can be
replaced by $\mathcal{R}_s$. Kinetic and strain energy changes
have the following asymptotic expressions:
\begin{equation}
    \Delta T\sim\ell^{2-2\beta}~~~~~\textrm{and}~~~~~\Delta U_e\sim\ell^{2-2\beta}.
\end{equation}
In addition to the kinetic and strain energies, there is another
important energy term, namely the surface energy. By assuming a
constant specific surface energy per unit of a fractal measure,
the surface energy that is required for the formation of a
self-affine fractal crack has the following asymptotic
behavior\footnote{For every short cracks surface energy is length
dependent \citep{Ippolito2006}. However, we are interested in
obtaining the limiting crack velocity that corresponds to crack
lengths much larger than any fracture quantuum and hence specific
surface energy is a material constant.}:
\begin{equation}\label{14}
   \Delta U_s\sim\ell^{\frac{1}{H}}.
\end{equation}

For quasi-static crack growth, Griffith's criterion can be written as $\Delta U_e+\Delta U_s=0$, while for dynamics crack growth it is written as $\Delta U_e+\Delta U_s+\Delta T=0$, where $\Delta U_e$ is the change of the strain energy in the body due to crack growth, $\Delta U_s$ is the required energy for the formation of the new fracture surfaces, and $\Delta T$ is the change of kinetic energy in the body. Using the equality of asymptotic expressions of energy terms the order of stress singularity can be obtained as
\begin{equation}\label{15}
    \ell^{2-2\beta}\sim\ell^{\frac{1}{H}}.
\end{equation}
Thus
\begin{equation}
    \qquad\beta=\frac{2H-1}{2H},
\end{equation}
which is identical to that of a stationary crack \citep{Mosolov1991,Gold1991,Gold1992,Balankin1997,Yavari2000,Yavari2002,Yavari2002a}. Therefore, the stress field has the following asymptotic form:
\begin{equation}
\sigma\sim r^{-\beta}\qquad \textrm{where} \qquad \beta=
\begin{cases}
\frac{2H-1}{2H}, &\text{$\frac{1}{2}<H<1$}\\
0,&\text{$0<H<\frac{1}{2}$}
\end{cases}.
\end{equation}

\section{Terminal velocity of rough crack growth in brittle materials}

From many studies of fracture surfaces formed in brittle materials, it is believed that the surfaces created by the process of dynamic fracture have a characteristic structure, referred to as \emph{mirror--mist--hackle} in the literature. This structure has been observed to occur in materials as diverse as glass and ceramics, noncrosslinked glassy polymers such as PMMA and crosslinked glassy polymers such as Homalite 100, polystyrene and epoxies
(for more details see \cite{Lawn1993,Gao1993,Hauch1998,Ravi1998,Fineberg1999} and references therein).

The prediction and measurement of the crack tip speed has received great attention from researchers in the field of dynamic fracture. As was mentioned earlier, \citet{Roberts1954} were the first to find a theoretical prediction for limiting crack tip velocity. Their calculations based on the Mott's extension of Griffith's criterion predicted crack tip speed of $0.38c_0$ (for $\nu=0.25$), where $c_0=\sqrt{E/\rho}$. \citet{Steverding1970} also predicted the terminal velocity by using an asymptotic solution and found it to be about $0.52c_R$. Many researchers argue that the maximum velocity attainable by any moving surface of discontinuity should be identified with the velocity of the Rayleigh surface waves $(c_R)$. \citet{Ravi1998} reached the following three major conclusions about the crack speed measurements: i) There is an upper limit to the speed with which dynamic cracks propagate. ii) This limiting crack speed is significantly lower than the Rayleigh surface wave speed of the material. iii) The limiting speed is not a fixed fraction of the Rayleigh wave speed; this fraction is material dependent. The data gathered by \citet{Fineberg1999} indicates that in amorphous materials such as PMMA and glass, the maximum observed velocity of crack propagation barely exceeds about $1/2$ of the predicted value.\footnote{Note that this is true only for isotropic materials. In an anisotropic body, terminal velocity can reach up to ninety percent of the Rayleigh wave speed. See the review by \citet{Fineberg1999} for more details. Here we restrict ourselves to crack propagation in an isotropic medium.} In the following we predict the terminal crack tip velocity using an asymptotic energy balance argument.

\begin{figure}
    \begin{center}
      \resizebox{90mm}{!}{\includegraphics{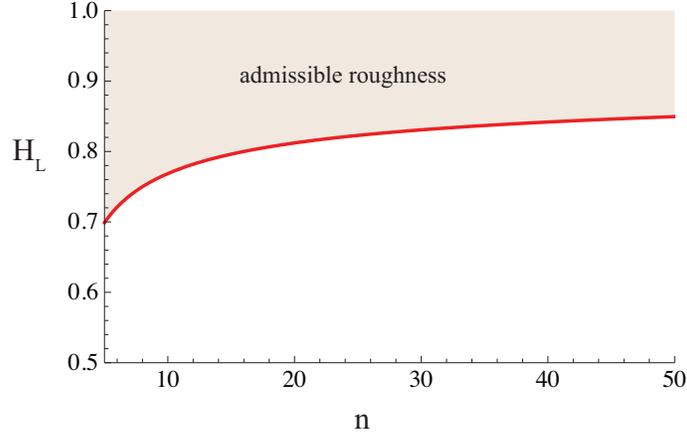}}
      \caption{Limiting roughness $H_L$ for different values of $n=\Delta L_{nominal}/a_0$.}
      \label{TerminalRoughness}
    \end{center}
\end{figure}

To illustrate the process of reaching a constant crack tip speed, suppose that an infinite domain $\mathcal{R}$ with an initially smooth crack is subjected to remote tensile stresses $\sigma_\infty$. The crack unloads some area of the body $\mathcal{R}_c$ that can be approximated by a disk $\mathcal{R}_s$ of radius $r_s$ \citep{Yavari2002}. To specify this circle we need to define a characteristic length for the problem. There is experimental evidence that the dynamic fracture processes approach a steady state and thus taking the crack length as a characteristic length will contradict a steady state condition.\footnote{In the case of a semi-infinite crack there is no characteristic length.} Therefore, we seek a new characteristic length in the problem. Here the fracture quantum is taken as the material characteristic length. Therefore, for a fractal crack the radius of the disk $\mathcal{R}_s$ is assumed to be proportional to $a_0$, i.e. $r_s\sim a_0$ from dimensional analysis arguments. More precisely, for a self-affine crack, the following relation holds for the radius of the dominance region of strain energy release:
\begin{equation}\label{16}
  r_s=a_0 \Phi(H),
\end{equation}
where $\Phi$ is a dimensionless function. We assume that the
crack is initially smooth ($H=1$). At time $t=0$ the strain
energy density reaches a critical value and suddenly the crack
starts to grow. As the nominal length of the crack increases, the
roughness of the fracture surfaces increases as well ($H$ decreases) due to the mirror-mist-hackle transition phenomenon. We postulate that crack surfaces reach a terminal roughness. It should be noted that a self-affine fractal model for
$H<\frac{1}{2}$ is a plane-filling set and hence
the limiting roughness lies in the range $\frac{1}{2}<H_L<1$. Let us justify our
postulate of reaching a terminal velocity using a crack branching argument. The required surface
energy for the formation of a self-affine fractal crack has the
asymptotic behavior $\ell^{\frac{1}{H}}$. To estimate the actual
growth of a self-affine fractal crack, suppose that the nominal
crack growth step is equal to $na_0$, where $a_0$ is fracture
quantum and $n>1$. According to fractal geometry
concepts the actual growth length $\Delta L$ of a self-affine fractal
crack with the nominal growth step size of $na_0$ has the asymptotic form $\Delta L\sim a_0n^{\frac{1}{H}}$. Therefore, the required
surface energy has the asymptotic behavior of the form
$n^{\frac{1}{H}}$. Now we argue that for some values of $H$ the
required surface energy for the formation of two new surfaces
(assuming that these new cracks are initially smooth) will become
smaller than the required surface energy for the continuation of
the single (roughened) crack.\footnote{\citet{Eshelby1971} suggested that a
crack would branch when the energy going into the creation of a
single propagating crack is enough to support two single cracks.
For more details see \citet{Fineberg1999}.} For roughness exponents smaller than this limiting roughness (denoted by
$H_L$) one can write:
\begin{equation}\label{16}
  (n)^{\frac{1}{H}}>2(n)~~~~~\textrm{for}~~H<H_L.
\end{equation}
Therefore, it is probable that by reaching $H_L$ the increase of
energy flow toward the crack tip causes branching of the crack. This
roughness limit can be estimated by solving $n^{\frac{1}{H}}=2n$
for different values of $n$.\footnote{Note that the equation that we need to solve is $k(H)n^{\frac{1}{H}}=2n$ for a function $k$ such that $k(1)=1$. As we do not have an explicit form for $k(H)$, we assume it is approximately equal to unity. If we assume other constant values for $k(H)$ the only change will be value of the approximate terminal roughness. It seems that the choice $k=1$ leads to a reasonable terminal roughness in agreement with experiments. We should also emphasize that the exact value of this terminal roughness will not change any of the subsequent results.} The results are presented in Fig.
\ref{TerminalRoughness}. Note that for each $n$ the acceptable
values of $H$ for having a single crack are $H\geq H_L$. There
are different arguments in the literature for explaining the
branching phenomena in fracture but many of them are not in
agreement with the experimental results.\footnote{Once a crack
bifurcates, single crack models are, of course, no longer valid.
Therefore, a theory describing a single crack can, at best,
provide a criterion for when crack branching occurs. A number of
such criteria for the onset of crack branching have been
proposed. The criterion due to \citet{Yoffe1951} and extremal
energy density criteria
\citep{Sih1973,Theocaris1985,Ramulu1986,Adda1996} all suffer from
the common problem that the velocities predicted for the onset of
branching are much higher than the observed velocities in the
experiments. Additional criteria such as postulating a critical value
of the stress intensity factor, have not been consistent with
experiments \citep{Ramulu1985,Arakawa1991} since measurements at
the point of branching show considerable variation of the stress
intensity factor $K_I$.} Note that in the
range $n\in[10,100]$, $H_L\approx 0.8$ is almost constant. For large
$n$, $H$ increases but very slowly. Note that as $n\rightarrow
\infty$, $\ln n =o(n^{\epsilon})~~~\forall~\epsilon>0$ \citep{Bleistein1986}. This means that $\ln
n$ increases indefinitely but very slowly as $n$ increases. Note
also that for short cracks surface energy depends on $n$ and
hence in Fig. \ref{TerminalRoughness}, $n\leq 5$ is not shown.

Since the pioneering work of \citet{Mandelbrot1984} understanding the morphology of fracture surfaces has been a very active field of research. Much
recent effort has been focused on characterizing fracture
surfaces in terms of a roughness exponent. There are many
researchers who argue that there exists a universal roughness for
fracture surfaces. Their studies on the fracture surfaces of
different materials indicate that for both quasi-static and
dynamic fracture a universal roughness exponent of approximate
value 0.8 can be obtained for values of $\xi$ (scale of
observation) greater than a material-dependent scale, $\xi_c$
\citep{Bouchaud1990,Bouchaud1997,Bouchaud2003,Maloy1992,Daguier1996,Daguier1997,Ponson2006}.
What we see from our simple crack branching argument is in agreement with these experimental studies.

By taking the nominal crack growth step to be $na_0$ the actual
crack growth length for a self-affine crack propagation can be estimated as
follows:
\begin{equation}
    \Delta L_{\text{actual}}=\sqrt{\delta x^2 +\delta y^2}\sim a_0\sqrt{n^2 +n^{\frac{2}{H}}}\sim a_0
    n^{\frac{1}{H}}~~~~~(n>1).
\end{equation}
Note that we assume that the crack lies in the $(x,y)$ plane with nominal growth in the $x$ direction. Let us define the nominal crack-tip velocity by $v_{\text{nominal}}=\frac{\Delta
L_{\text{nominal}}}{\Delta t}=\frac{na_0}{\Delta t}$ and the actual
velocity by $v_{\text{actual}}=\frac{n^\frac{1}{H} a_0}{\Delta t}$, where
$\Delta t$ is the required time for growth of the fractal
crack by the nominal amount of $n a_0$. Therefore, the following relation
holds between nominal and actual velocities of the crack tip:
\begin{equation}\label{Nominal-actual-relation}
    v_{\text{actual}}\sim n^{\frac{1}{H}-1}v_{\text{nominal}}.
\end{equation}
When the crack roughness reaches its limiting value $H_L=H_L(n)$,
$n^{\frac{1}{H_L(n)}-1}=2$ and hence $v_{\text{actual}}\sim
2v_{\text{nominal}}$. Now if the limit of $v_{\text{actual}}$ is
$c_R$, we see that limit of $v_{\text{nominal}}$ is about $\frac{1}{2}c_R$
from this rough estimate. \citet{Gao1993} used a wavy-crack model
and observed that depending on roughness local crack tip velocity
can be as large as twice the apparent crack tip speed. He also observed that when
apparent crack speed is $\frac{1}{2}c_R$ dynamic energy release rate
is maximized.

Reaching the terminal roughness exponent
$H_L$ has another important consequence; after reaching $H_L$ the
radius of the dominance zone of strain energy release will
remain unchanged, i.e. $r_s$ reaches a
constant value. Now we are back to the main problem of having a
terminal velocity for crack propagation in brittle materials. As we concluded earlier
from the experimental observations, almost all the released
strain energy will be converted into kinetic energy. Therefore, we are
concerned with the changes of strain and kinetic energies, i.e.
$\Delta U_e$ and $\Delta T$. The dominant change of strain energy can be written as follows:
\begin{equation}\label{strain-energy-variation}
  \Delta
  U_e=\int_{\mathcal{R}_s}\frac{1}{2}\sigma_{ij}\epsilon_{ij}dA.
\end{equation}
Similar to the case of dynamic fracture of a smooth crack, for a fractal crack the following stress and strain fields at the moving crack tip are assumed:
\begin{equation}
    \sigma_{ij}(r,\theta,v) = K_I^f(v)r^{-\beta}\Sigma_{ij}(\theta,v)~~~~~\text{and}~~~~~
  \epsilon_{ij}(r,\theta,v) = K_I^f(v)r^{-\beta}C_{ijkl}\Sigma_{kl}(\theta,v).
\end{equation}
Substituting the above asymptotic fields into Eq.(\ref{strain-energy-variation}), we obtain:
\begin{equation}\label{18}
  \Delta
  U_e=\int_{\mathcal{R}_s}\frac{1}{2}\left[{K_I^f(v)}\right]^2r^{-2\beta}\Sigma_{ij}(\theta,v)C_{ijkl}\Sigma_{kl}(\theta,v)dA.
\end{equation}
In the case of a mode I smooth crack, stress field explicitly depends on the instantaneous crack tip speed $v(t)$ \citep{Freund1998}. An immediate consequence of this is that the near tip field for the time-dependent motion is identical to that for steady state crack growth in the same material up to a time-dependent proportionality constant. This was demonstrated by \citet{Freund1974}, \citet{Nilsson1974}, and \citet{{Achenbach1975}}, each of whom compared asymptotic solutions for nonuniform crack growth with earlier results based on the assumption of steady growth obtained by \citet{Cotterell1964}, \citet{Rice1968}, and \citet{Sih1970}.

\citet{Freund1972a,Freund1972b} suggested an indirect method for determining the dynamic stress intensity factor for mode I crack propagation with nonuniform crack growth and a general loading condition. The result was that the dynamic stress intensity factor for arbitrary motion of the crack tip is proportional to the corresponding quasi-static stress intensity factor with a universal proportionality constant, i.e. $K_I(a,v)=k(v)K_I(a,0)$, where $a$ is the instantaneous crack characteristic length, and
\begin{equation}
  k(v)=\frac{1-v/c_R}{S_{+}(1/v)\sqrt{1-v/c_d}},
\end{equation}
where $c_R, c_d$, and $v$ are the Rayleigh wave speed, dilatational wave speed, and the crack tip velocity, respectively, and $S_{+}(1/v)$ is close to unity. Therefore, $k(v)$ can be approximated by
\begin{equation}
  k(v)=\frac{1-v/c_R}{\sqrt{1-\left(v/c_R\right)/\xi}},
\end{equation}
where $\xi=c_d/c_R$.\footnote{Note that for plane strain
$c_d/c_R\in[1.635,\infty]$, and for plane stress
$c_d/c_R\in[1.635,2.145]$.} We assume that the above result
holds for a fractal crack, i.e.\footnote{\citet{Xie1995} made a
similar assumption for each prefractal crack trajectory in his
fractal model.}
\begin{equation}
  K_I^f(\textit{a},v)=k(v)K_I^f(a,0).
\end{equation}

\begin{figure}
    \begin{center}
      \resizebox{100mm}{!}{\includegraphics{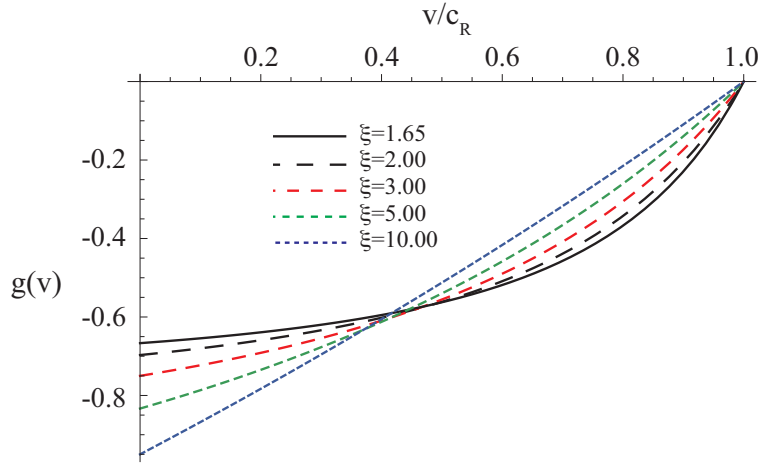}}
      \caption{Schematic plot of the strain energy variation function, $g(v)=k(v)k'(v)$ versus normalized crack tip speed $v/c_R$.}
      \label{StrainEnergy}
    \end{center}
\end{figure}

The quasi-static stress intensity factor for a fractal crack of projected length $2a$ subjected to uniform far-field stress $\sigma^\infty$ is
$K_{I0}^f=\psi(H)\sigma^\infty \sqrt{\pi a^{(2H-1)/H}}$ \citep{Yavari2002a,Wnuk2003}. Thus, for very long cracks (intermediate crack growth regime) the rate of change of stress intensity factor becomes vanishingly small, i.e.
\begin{equation}
      \frac{\partial K_{I0}^f}{\partial a}\sim 0~~~\textrm{as}~~~a\rightarrow\infty.
\end{equation}
Now, we can use the above arguments in the calculation of the strain energy release. If roughness reaches its terminal value $H_L$, the size of the dominance zone of strain energy release $\mathcal{R}_s$ will became approximately constant and as a result the order of stress singularity will remain unchanged. In addition to this, increasing the nominal crack length to large values the change in the stress intensity factor becomes vanishingly small. In other words, the stress intensity factor becomes approximately constant. Under these conditions, strain energy will change only due to the change of the nominal velocity of the crack tip. Strain energy in the disk $\mathcal{R}_s$ is $(H \to H_L, a\gg 1)$:
\begin{gather}
    \nonumber
    \Delta U_e(v)\sim \frac{1}{2}k(v)^2\int_{\mathcal{R}_s(H_L)}\left[K_{I0}^f\right]^2r^{-\frac{2H_L-1}{H_L}}\Sigma_{ij}(\theta,v)C_{ijkl}\Sigma_{kl}(\theta,v)dA\\
     ~~~~~=\frac{1}{2}\left[K_{I0}^f\right]^2k(v)^2H_L~r_s^{\frac{1}{H_L}}\int_{0}^{2\pi}\Sigma_{ij}(\theta,v)C_{ijkl}\Sigma_{kl}(\theta,v)d\theta,
\end{gather}
where $v=v_{\text{nominal}}$ for short. It can be shown that the effect of infinitesimal changes of crack tip velocity $(\delta v)$ on the angular variation of the stress field is negligible and hence, the change of strain energy in the disk $\mathcal{R}_s$ due to the change of crack tip velocity can be simplified to read\footnote{Note that $\Delta(*)$ denotes change in the quantity $*$ due to crack growth (change in the crack length) while $\delta(\star)$ denotes change in the quantity $\star$ due to change in the crack speed.}
\begin{equation}
    \delta \Delta U_e = \frac{1}{2}\left[k(v+\delta v)^2 C_{U_e}(H_L,\sigma^\infty)-k(v)^2 C_{U_e}(H_L,\sigma^\infty)\right]
    \sim  C_{U_e}(H_L,\sigma^\infty)k(v)k'(v)\delta v,
\end{equation}
where
\begin{equation}
    C_{U_e}(H_L,\sigma^\infty)=\left[K_{I0}^f\right]^2H_L~r_s^{\frac{1}{H_L}}\int_{0}^{2\pi}\Sigma_{ij}(\theta,v)C_{ijkl}\Sigma_{kl}(\theta,v)d\theta\geq 0.
\end{equation}
Fig. \ref{StrainEnergy} schematically shows the behavior of the function $g(v)=k(v)k'(v)$. Note that the above integral is positive and hence strain energy change is always negative; for all values of $c_d/c_R$ and $v/c_R$ strain energy is released as expected.

Now, we assume that the change of kinetic energy due to the
released strain energy is dominant in a disk $\mathcal{R}_k$ with
radius $r_k$. Similar to the case of determining radius of
dominance zone of strain energy release, the fracture quantum
($a_0$) is the characteristic length in the problem. Assuming
that $r_k=r_k(a_0,H,c_R,v_{\text{nominal}})$, dimensional
analysis tells us that:
\begin{equation}
    r_k=a_0~\Theta\!\left(H,\frac{v_{\text{nominal}}}{c_R}\right),
\end{equation}
where $c_R$ is Rayleigh wave speed, and $\Theta$ is a dimensionless function. Now we have the following relation for the amount of kinetic energy change in the disk $\mathcal{R}_k$:
\begin{equation}\label{17}
    \Delta T=\int_{\mathcal{R}_k}\frac{1}{2}\rho\dot{u}_i\dot{u}_i\,dA,
\end{equation}
where $\dot{u}_i$ is the velocity components of material particles. At the radial distance $r$ from the crack tip, the particle velocity depends on the crack tip velocity and the strain at the particle position. The velocity of particles in the dominance zone of kinetic energy, $\mathcal{R}_k$, is also a function of the angular position of particles and the ratio of the crack tip speed and a characteristic wave speed. If we assume that for the limiting roughness $H_L$ the effect of the change of the nominal crack tip speed on the size of the dominance zone is negligible, the dominance zone of the kinetic energy change is fixed, and once again these conditions dictate that kinetic energy can be written as:
\begin{equation}
   \Delta T=k(v)^2 v^2
    \int_{\mathcal{R}_k}\frac{1}{2}\rho\left[\frac{K_{I0}^f}{r^{\beta}E}\Psi_{i}(\theta,v)\right]^2 dA,
\end{equation}
where $v$ is short for $v_{\text{nominal}}$. We know that the effect of infinitesimal changes of crack tip velocity $(\delta v)$ on the angular variation of the particle velocity field is negligible, and hence the change of kinetic energy due to the change of crack tip velocity can be written as
\begin{eqnarray}
  \delta \Delta T &=& \frac{1}{2}\left[k(v+\delta v)^2 (v+\delta v)^2
    C_T(H_L,\sigma^\infty)-k(v)^2 v^2
    C_T(H_L,\sigma^\infty)\right] \nonumber \\
    & \sim & C_T(H_L,\sigma^\infty)\left[k(v)k'(v)v^2+k(v)^2 v\right]
          \delta v,
  \end{eqnarray}
where
\begin{equation}
    C_T(H_L,\sigma^\infty)=\rho\left[\frac{K_{I0}^f}{E}\right]^2\!\!H_L~r_k^{\frac{1}{H_L}}\!\int_{0}^{2\pi}\Psi_{i}(\theta,v)\Psi_{i}(\theta,v)d\theta\geq 0.
\end{equation}
\begin{figure}
    \begin{center}
      \resizebox{170mm}{!}{\includegraphics{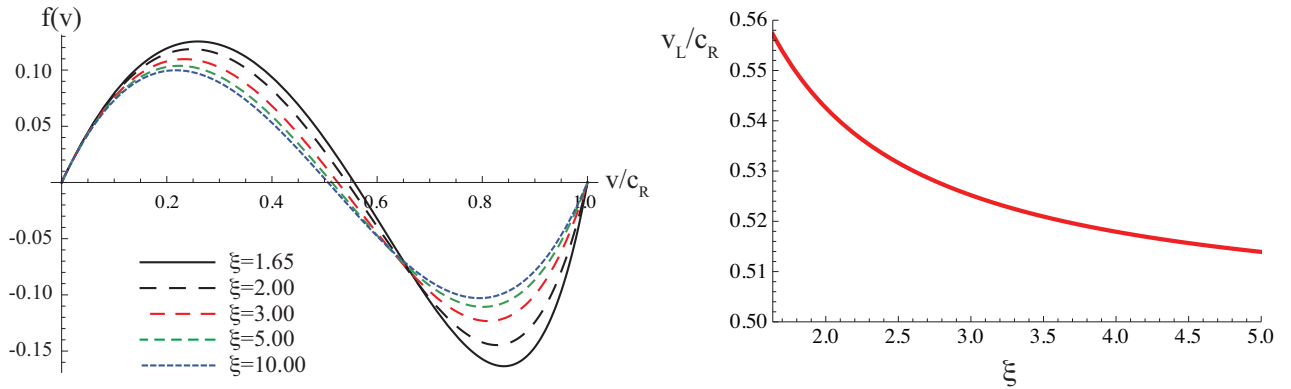}}
      \caption{a) Schematic plot of the kinetic energy variation function, $f(v)=k(v)k'(v)v^2+k(v)^2 v$ versus normalized crack tip speed $v/c_R$. b) Terminal velocity versus $\xi=c_d/c_R$.}
      \label{TerminalVelocity}
    \end{center}
  \end{figure}
Fig. \ref{TerminalVelocity}(a) schematically shows the function
$f(v)=k(v)k'(v)v^2+k(v)^2 v$ for different values of
$\xi=c_d/c_R$. Note that in plane strain
$c_d/c_R\in[1.635,\infty]$, and for plane stress
$c_d/c_R\in[1.635,2.145]$ and hence the appropriate range of
$c_d/c_R$ should be considered for each case.

The schematic graph of the function $k(v)k'(v)v^2+k(v)^2 v$ in
Fig. \ref{TerminalVelocity}(a) shows an interesting phenomenon:
After reaching some value of crack tip speed $(v_L)$ increasing
$v$ the change of kinetic energy becomes negative. Because the
particle velocity inside the dominance zone of kinetic energy is
proportional to the crack tip speed increasing the crack tip
speed the kinetic energy change must be positive, and therefore
the crack can not pass this limiting speed $(v_L)$. In other
words, when the crack tip speed increases the kinetic energy of
the region around the crack tip must increase as well, i.e. kinetic energy must be an increasing function of the crack tip speed. In terms of the equivalent crack tip mass introduced by
\citet{Chekunaev2008,Chekunaev2009}, our argument is equivalent to
saying that the equivalent mass must always be positive (note that their equivalent mass is independent of velocity), which is the case as was mentioned in \S1.

The predicted limiting speed for Glass, Homalite-100, PMMA,  K5 (Glass), K6 (Glass), and SF6 (Glass) are as follows:
\begin{itemize}
    \item [i)] Glass \hskip 0.53 in ($\nu=0.220$):\hskip 0.15 in   $\tilde{v}_L=0.321c_0,~~ \hat{v}_L=0.320c_0$,
    \item [ii)] Homalite-100 \hskip 0.05in ($\nu=0.310$):\hskip 0.15in $\tilde{v}_L=0.311c_0,~~ \hat{v}_L=0.310c_0$,
    \item [iii)] PMMA \hskip 0.39 in ($\nu=0.350$):\hskip 0.15 in $\tilde{v}_L=0.305c_0,~~ \hat{v}_L=0.306c_0$,
     \item [iv)] K5 (Glass) \hskip 0.2in ($\nu=0.227$):\hskip 0.15in $\tilde{v}_L=0.320c_0,~~ \hat{v}_L=0.319c_0$,
     \item [v)] K6 (Glass) \hskip 0.2in ($\nu=0.231$):\hskip 0.15in $\tilde{v}_L=0.320c_0,~~ \hat{v}_L=0.318c_0$,
     \item [vi)] SF6 (Glass) \hskip 0.15in ($\nu=0.248$):\hskip 0.15in $\tilde{v}_L=0.318c_0,~~ \hat{v}_L=0.317c_0$,
\end{itemize}
where $\tilde{v}_L=v_L^{\text{plane~strain}}$ and $\hat{v}_L=v_L^{\text{plane~stress}}$.
\begin{table}[h]
\begin{center}
    \begin{tabular}{l l c c c}
\hline
Material & Author &  $v_L/c_0$ & $v_L/c_R$ \\
\hline
Glass   & \citet{Schardin1938}&  0.30 & 0.52 \\
$\nu=0.22$        & \citet{Edgerton1941}&  0.28 & 0.47 \\
        & \citet{Bowden1967} &  0.29 & 0.51\\
        & \citet{Anthony1970}&  0.39 & 0.66 \\
\hline
PMMA  & \citet{Dulaney1960}& 0.36 & 0.62 \\
$\nu=0.35$      & \citet{Cotterell1965} & 0.33 & 0.58\\
      & \citet{Paxson1973}&  0.36 & 0.62 \\
      & \citet{Fineberg1992}& & 0.58-0.62 \\
\hline
Homalite-100 & \citet{Beebe1966}  & 0.19 & 0.33\\
$\nu=0.31$   & \citet{Kobayashi1978} & 0.22 & 0.37 \\
             & \citet{Dally1979}&  0.24 & 0.38 \\
             & \citet{Ravi1984a,Ravi1984b,Ravi1984c,Ravi1984d}&    & 0.45 \\
             & \citet{Hauch1998}&    & 0.37 \\
\hline
K5 (Glass)  &\citet{Senf1994} & 0.29-0.3& 0.5-0.52\\
$\nu=0.227$& & &\\
\hline
K6 (Glass)  &\citet{Senf1994} & 0.27-0.3& 0.47-0.51\\
$\nu=0.231$& & &\\
\hline
SF6 (Glass)   &\citet{Senf1994} & 0.2-0.23& 0.34-0.4\\
$\nu=0.248$& & &\\
\hline
\end{tabular}\label{TableVT}
\caption{Experimental values of limiting crack speeds for brittle amorphous materials.
$c_0,c_R$, and $\nu$ are the longitudinal sound wave speed, Rayleigh wave speed, and Poisson's ratio, respectively.}
\end{center}
\end{table}
It is seen from Table 1. that there is some scatter in the
experimental data and this makes any comparison with experimental
data difficult. However, we see that our estimates are close to
the experimental data.  It should be noted that what we have
obtained for terminal velocity is an upper bound.\footnote{Note that any velocity larger than our calculated $v_L$ leads to a decreasing kinetic energy change that is not physical. However, this does not mean that a lower terminal velocity is not possible. In other words, what we calculate is an upper bound to terminal velocity.} Interestingly,
all the experimentally measured velocities (except PMMA) are
smaller than our prediction of the terminal velocity.

\begin{figure}
    \begin{center}
      \resizebox{165mm}{!}{\includegraphics{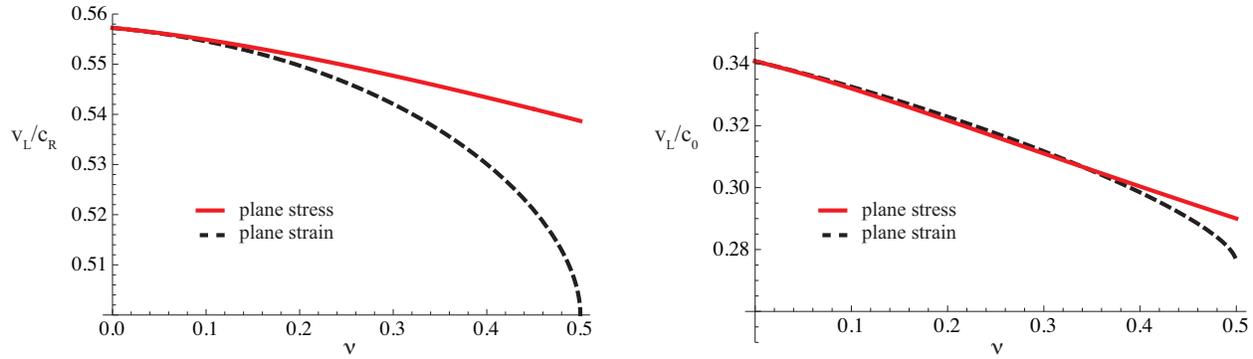}}
      \caption{Normalized limiting crack speed versus Poisson's ratio in plane stress and plane strain.}
      \label{Poisson}
    \end{center}
\end{figure}

The relations between various wave speeds and $c_0$ are:
$c_d=\sqrt{\frac{1-\nu}{(1+\nu)(1-2\nu)}}c_0$ (plane strain
dilatational wave speed), $c_d^p=\frac{c_0}{\sqrt{(1-\nu^2)}}$
(plane stress dilatational wave speed),
$c_s=\frac{c_0}{\sqrt{2(1+\nu)}}$ (shear wave speed),
$c_R=c_s\bigl(1-\frac{0.135}{3-4k^2}\bigl)$ (Rayleigh wave
speed), where $k^2=\frac{1-2\nu}{2(1-\nu)}$ for plane strain and
$k^2=\frac{1-\nu}{2}$ for plane stress. In Fig. \ref{Poisson} the
normalized limiting velocities $v_L/c_R$ and $v_L/c_0$ are plotted for different
values of Poisson's ratio and for both cases of plane stress and
plane strain. The calculations show that depending on the
Poisson's ratio the limiting velocities are in the range
$0.276c_0-0.341c_0$ for plane strain and in the range
$0.290c_0-0.341c_0$ for plane stress. 

\section{Conclusions}

In this paper, we first obtained the asymptotic stress field
around the tip of a dynamically propagating self-affine fractal
crack. We then showed that there is always a lower bound to
roughness exponent. We next looked at crack propagation and the
asymptotic behaviors of kinetic and strain energy changes due to
crack growth. We obtained an upper bound for terminal velocity by
postulating that the kinetic energy change must be a
monotonically increasing function of nominal crack tip speed. We
predicted a material-dependent terminal velocity in the range
$[0.500c_R,0.557c_R]$ and $[0.539c_R,0.557c_R]$ for plane stress and plane strain, respectively. We should
emphasize that our asymptotic analysis only gives an estimate of
terminal velocity. It was observed that for several amorphous
brittle materials our predicted terminal velocities are in good
agreement with the experimental data. In summary, our main
results are: i) Fractal cracks tend to reach an approximately
constant terminal roughness close to $H_L=0.8$. ii) There is a
terminal velocity of crack propagation lower than the Rayleigh
wave speed. iii) The terminal velocity is a material-dependent
fraction of the corresponding Rayleigh wave speed. iv) This
material-dependent fraction only depends on the Poisson's ratio.



\end{document}